# Radioactive $^{26}$Al and massive stars in the Galaxy

Roland Diehl[1], Hubert Halloin[1], Karsten Kretschmer[1], Giselher G. Lichti[1], Volker Schönfelder[1], Andrew W.Strong[1], Andreas von Kienlin[1], Wei Wang[1], Pierre Jean[2], Jürgen Knödlseder[2], Jean-Pierre Roques[2], Georg Weidenspointner[2], Stephane Schanne[3], Dieter H. Hartmann[4], Christoph Winkler[5], and Cornelia Wunderer[6]

[1]Max-Planck-Institut für extraterrestrische Physik, D-85748 Garching, Germany
[2]Centre d'Etude Spatiale des Rayonnements and Université Paul Sabatier, 31028 Toulouse, France
[3]DSM/DAPNIA/Service d'Astrophysique, CEA Saclay, 91191 Gif-Sur-Yvette, France
[4]Clemson University, Clemson, SC 29634-0978, USA
[5]ESA/ESTEC, SCI-SD 2201 AZ Noordwijk, The Netherlands
[6]Space Sciences Lab., Berkeley, CA 94720, USA

**Gamma-rays from radioactive $^{26}$Al (half life ~7.2 $10^5$ yr) provide a 'snapshot' view of ongoing nucleosynthesis in the Galaxy[1]. The Galaxy is relatively transparent to such gamma-rays, and emission has been found concentrated along the plane of the Galaxy[2]. This led to the conclusion[1] that massive stars throughout the Galaxy dominate the production of $^{26}$Al. On the other hand, meteoritic data show locally-produced $^{26}$Al, perhaps from spallation reactions in the protosolar disk[4,5,6]. Furthermore, prominent gamma-ray emission from the Cygnus region[2,3] suggests that a substantial fraction of Galactic $^{26}$Al could originate in localized star-forming regions. Here we report high spectral resolution measurements of $^{26}$Al emission at 1808.65 keV, which demonstrate that the $^{26}$Al source regions corotate with the Galaxy, supporting its Galaxy-wide origin. We determine a present-day equilibrium mass of 2.8 (±0.8) $M_\odot$ of $^{26}$Al. We use this to estimate that the frequency of core collapse (i.e. type Ib/c and type II) supernovae to be 1.9(± 1.1) events per century.**

Excess $^{26}$Mg found in meteorites shows that the hot disk-accretion phase of the presolar nebula apparently was characterized[4,5] by an amount of radioactive $^{26}$Al (relative to the stable $^{27}$Al isotope) with a rather well-determined $^{26}$Al/$^{27}$Al ratio of ~4.5 $10^{-5}$. This is surprising, given that $^{26}$Al decays within ~1Myr: The time it takes for a parental molecular cloud after decoupling from nucleosynthetically-enriched interstellar gas to form protostellar disks is much longer[7]. Therefore, the meteoritic determinations of the $^{26}$Al/$^{27}$Al ratio have been interpreted as an *in situ* $^{26}$Al enrichment of the young solar nebula[8], by either a nearby supernova or a AGB star event injecting fresh nucleosynthesis products at a "last moment", or by enhanced cosmic-ray nucleosynthesis in the magnetically-active early Sun with its accretion disk[6]. The *mean* $^{26}$Al content of the interstellar medium in the Galaxy would therefore decouple from the solar value. Observation of 1808.65 keV gamma-rays from the decay of radioactive $^{26}$Al in the interstellar medium, however, demonstrated that $^{26}$Al nucleosynthesis does occur in the present Galaxy. The somewhat irregular distribution[2,3] of $^{26}$Al emission seen along the plane of the Galaxy provided a main argument for the idea that massive stars dominate the production of $^{26}$Al[1]. Massive stars preferentially form in clusters; some of the nearby massive-star regions appear prominent in $^{26}$Al emission (e.g. in the Cygnus region), while others do not. Since the massive star census in the Galaxy is well known only out to distances of a few kpc, and many regions of the Galaxy are occulted for direct measurements, one is left with considerable uncertainty about a Galaxy-wide interpretation of the gamma-ray measurements, versus the possibility of localized efficient $^{26}$Al-producing regions. The total amount of $^{26}$Al in the Galaxy, and hence the mean interstellar $^{26}$Al/$^{27}$Al ratio, is thus rather uncertain.

If $^{26}$Al sources are indeed distributed throughout the Galaxy, Galactic rotation will cause Doppler shifts of the gamma-ray line energy, depending on the location of the source region within the Galaxy. Offsets in the line energy range up to 0.25 keV, and are particularly pronounced towards longitudes around ±30° (see Fig. 2)[9]. From 1.5 years of data from our Ge spectrometer on the INTEGRAL gamma-ray observatory of ESA (launched in October 2002), we derived new spectra of celestial $^{26}$Al emission (see Online Supplementary Information for details of observations and analysis method). From the inner Galaxy, we obtain an $^{26}$Al measurement at 16σ above background, and significantly (>3σ) detect the $^{26}$Al line emission in six 0.5 keV-wide energy bins across the line center – this allows an unprecedented investigation of $^{26}$Al gamma-ray line parameters (Fig. 1). The observed line shape is found to approach the one expected from instrumental resolution. Hence the line width of celestial emission, which would be observable as additional broadening, is rather small. The high spectral precision of our instrument can now be combined with its imaging capability to search for signatures of Galactic rotation, i.e. to determine $^{26}$Al line centre energies for different regions along the Galactic plane.

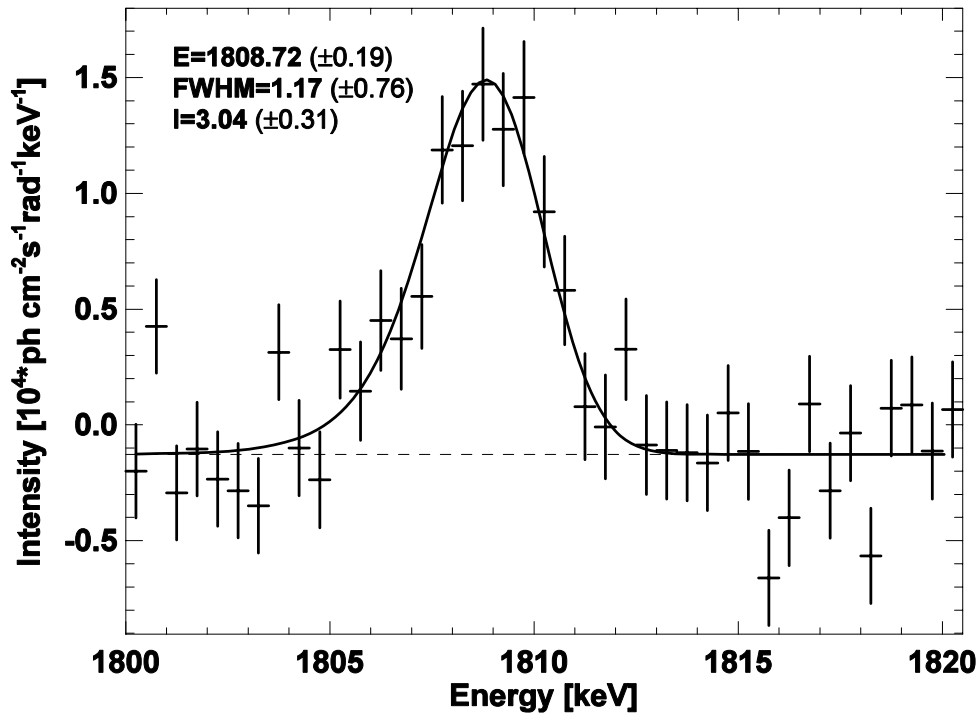

**Fig.1:** *Measurement of the $^{26}$Al line from the inner Galaxy region with SPI/INTEGRAL. Our INTEGRAL sky exposure is fairly symmetric around the centre of the galaxy and extends over the full Galactic plane, though emphasizing the range of ±45° longitude and ±15° latitude (see Online Supporting Information). This spectrum (shown with s.d.error bars) was derived from sky model fitting to the set of 19 Ge detector count spectra for each of the 7130 spacecraft pointings, using the COMPTEL $^{26}$Al image as a model for the spatial disctribution of emission, and background modeled from auxilliary measurements. The fit (solid line) combines the instrumental resolution, which is derived by accounting for degradation from cosmic-ray degradation irradiation and for annealings during the time of our measurement, with a gaussian for the intrinsic, astrophysical $^{26}$Al line width. For this integrated result from the inner Galaxy, the line centre is determined to be at 1808.72 (±0.2(stat)±0.1(syst)) keV, well within the laboratory value for the $^{26}$Al line of 1808.65 (7) keV. The integrated intensity from the inner region of the Galaxy is determined to 3.3 (±0.4) $10^{-4}$ ph cm$^{-2}$ s$^{-1}$ rad$^{-1}$, averaging over this and other plausible spatial-distribution models. (see Online Supporting Information for method and details)*

Our method to derive spectra must adopt a model for the distribution of emission over the sky (see Online Supporting Information). When we split such a model in the interesting region –40°…+40° into three longitude segments through cuts at –10° and 10°, and determine $^{26}$Al line energies for these regions, we systematically find that the $^{26}$Al line energy clearly falls above the laboratory energy east of the Galactic center, and slightly below towards the west (see Fig. 2 and Online Supporting Information). Such a signature is expected from Galactic rotation. This strongly supports the view that the observed $^{26}$Al source regions are located in the inner region of the Galaxy, rather than in localized foreground regions, as they appear to move along with the global Galactic interstellar medium. This affirms a Galaxy-wide interpretation of the $^{26}$Al gamma-ray measurement, which indirectly had been argued before on grounds of correlating the $^{26}$Al gamma-ray image with different tracers of candidate sources[1,10].

The total $^{26}$Al gamma-ray flux which we obtain is 3.3(±0.4) 10$^{-4}$ ph cm$^{-2}$s$^{-1}$ (this value is conventionally quoted for the inner Galaxy region, -30°<l<30°; -10°<b<10°); the flux varies by ~4% when we use a range of models, which we consider as plausible tracers of $^{26}$Al sources[1,10] (see Online Supporting Information). This can be converted to an equilibrium mass produced by ongoing nucleosynthesis throughout the Galaxy in steady state, once the 3D spatial distribution is known. Our best 3D model is based on free electrons liberated by ionising radiation from massive stars – these electrons can be measured at radio wavelengths, the results of which have been translated into such a model[11,12]. Using this and alternative plausible models[13] (see Online Supporting Information), we infer a mass of 2.8 M$_\odot$ of $^{26}$Al in the entire Galaxy. We estimate this value to be uncertain by ±0.8 M$_\odot$, from the statistical uncertainty of our measurements and the spatial model uncertainty. The $^{26}$Al/$^{27}$Al ratio implied[13,14] for the average interstellar medium is 8.4 10$^{-6}$ (assuming an interstellar gas mass of 4.95 10$^9$ M$_\odot$, and a abundance by number of log N($^{27}$Al) = 6.4, on a scale given by log N(H) = 12 ), about one order of magnitude lower than the solar-nebula value.

In conjunction with stellar yields and a distribution function of stellar birth masses this provides an independent estimate of the star-formation rate in the Galaxy. While we know reasonably well star formation rates for external galaxies and specific regions in the solar neighbourhood, the star formation rate of the Galaxy as a whole is much less certain due to occultation from interstellar clouds or other biases. Values based on optical to infrared tracers range from 0.8 to 13 M$_\odot$ per year[15,16,17,18] (see Online Supporting Information). The gamma-ray technique has the advantage in that it measures the rate in penetrating radiation over the full Galaxy, and averages over a time scale associated more closely with *one (current)* generation of massive stars (1 Myr).

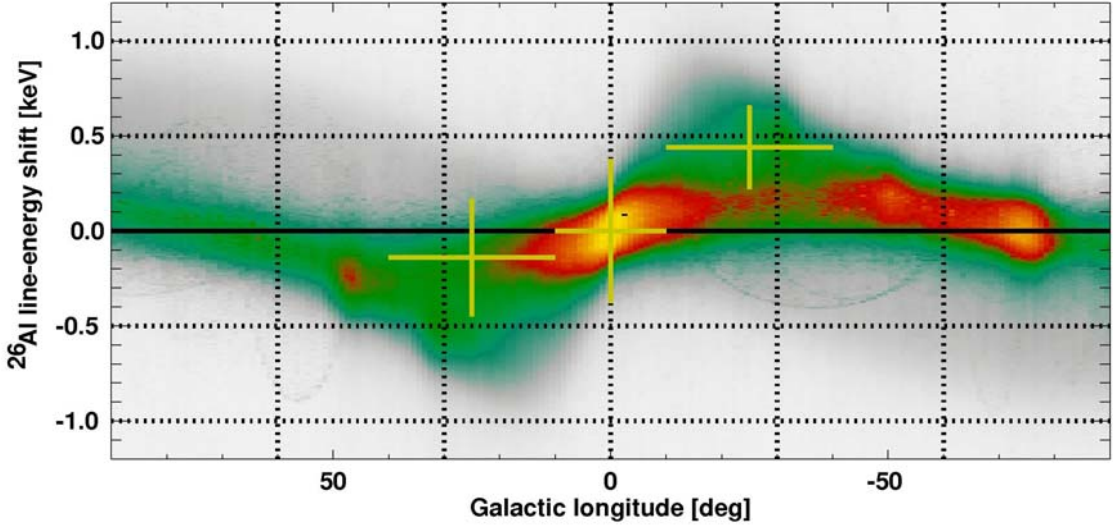

**Fig.2:** *Line position shifts with viewing directions along the inner Galaxy. Galactic rotation will shift the observed $^{26}$Al line energy due to the Doppler effect, to appear blue-shifted at negative longitudes and red-shifted at positive longitudes. Expectations (grey scale) have been modeled from the Galactic rotation curve and a 3-dimensional distribution of $^{26}$Al sources. From such models, we typically expect Doppler shifts of 0.25 keV, varying by ~0.05 keV with assumptions about inner-Galaxy rotation and spatial source distribution (see Online Supporting Information), for the integrated longitude ranges 40°..10° and −10°…-40°, respectively. As one of our models, we show here (grey scale) the one based on free electrons in the insterstellar medium[11,12], with an exponential distribution perpendicular to the Galactic plane (scale height 180 pc); for our longitude intervals 40°..10° and −10°…-40°, this predicts integrated Doppler shifts of -0.22 keV and +0.24 keV, respectively. Our observations (s.d. error bars shown) are consistent with the Galactic-rotation explanation at the 94% probability level. (see Online Supporting Information for method and details).*

The signature of Galactic rotation in the $^{26}$Al gamma-ray line reaffirms that large-scale distributions for tracers of $^{26}$Al sources can be applied to obtain a census of massive stars in the Galaxy.

Theoretical nucleosynthesis yields have been derived for massive stars, which are presumed to dominate the Galactic $^{26}$Al budget, specifically for core-collapse supernovae and for the preceding Wolf-Rayet phases. Such models have been shown to match the chemical history of the Galaxy as reflected in the abundances of chemical elements to within a factor of two[19], which is an impressive success. $^{26}$Al yields (wind-phase and explosive yields) from recent models of several independent research groups[20,21,22] converge within about 50% over the full mass range (see Online Supporting Information). Yields are moderated by the steep initial mass function (IMF), $\xi \sim m^{-\alpha}$, in our relevant mass range ~10-120 M$_\odot$. We use the Scalo IMF ($\xi \sim m^{-2.7}$) for this higher-mass range, supported by a wide range of astronomical constraints[23], to obtain an average ejected mass per massive star of $^{26}$Al of

$Y_{26}=1.4 \times 10^{-4}$ $M_{\odot}$. With our measured amount of $^{26}$Al, this IMF-weighted $^{26}$Al yield implies a rate of core-collapse supernovae in the Galaxy, averaged over the radioactive lifetime of $^{26}$Al (1.04 My); this represents the *current* core-collapse supernova rate: Evolutionary times for star clusters are ~10-100 My, while formation times of star clusters from Giant Molecular clouds are ~100 My, and the Galaxy's age is ~12 Gy. (See Online Supporting Information for the method of calculation and discussion of uncertainties). Our $^{26}$Al measure implies a rate of (1.9 ± 1.1) core-collapses per century, corresponding to a star formation rate of ~4 $M_{\odot}$ $y^{-1}$, or a stellar production rate of ~7.5 stars per year. Our high-resolution spectroscopy of $^{26}$Al with INTEGRAL/SPI data shows that the Galaxy produces stars at a moderate rate, typical for spiral galaxies of similar type and luminosity.

*Acknowledgements.* This paper is based on observations with INTEGRAL, an ESA project with instruments and a science data center funded by ESA member states (especially the PI countries: Denmark, France, Germany, Italy, Switzerland, Spain), Czech Republic and Poland, and with the participation of Russia and the USA. The SPI project has been completed under the responsibility and leadership of CNES/France. The SPI anticoincidence system is supported by the German government. We are grateful to ASI, CEA, CNES, DLR, ESA, INTA, NASA and OSTC for support. We are grateful to Alessandro Chieffi, Nikos Prantzos, and Stan Woosley for discussions of theoretical nucleosynthesis yields.

## References

1. Prantzos, N. & Diehl, R., Radioactive $^{26}$Al in the Galaxy: Observations versus theory. *Phys. Rep.,* **267, 1**,1 (1996)
2. Diehl, R., Dupraz, C., Bennett, K. *et al.,* COMPTEL observations of Galactic $^{26}$Al. *Astron. Astroph.*, **298**, 445-460 (1995)
3. Plüschke, S., Diehl, R., Schönfelder, V., *et al.,* The COMPTEL 1.809 MeV Survey. *ESA-SP,* **459**, 55-58 (2001)
4. MacPherson, G.J., Davis, A.M., Zinner, E.K., Distribution of $^{26}$Al in the early solar system. *Meteoritics*, **30**, 365 (1995)
5. Young, E.D., Simon, J.I., Galy, A., et al., Supra-Canonical $^{26}$Al/$^{27}$Al and the Residence Time of CAIs in the Solar Protoplanetary Disk, *Science*, **308**, 223-227 (2005)
6. Lee, T., Shu, F.H., Shang, H., et al., Protostellar cosmic rays and extinct radioactivities in meteorites, *Astrophys. J.,* **506**, 898-912 (1998)
7. Ward-Thompson, D., Isolated Star Formation: From Cloud Formation to Core Collapse, *Science*, **295**, 76-81 (2002)
8. Meyer, B.S., Clayton, D.D., Short-Lived Radioactivities and the Birth of the Sun. *SSRev*, **92**, 133-152 (2000)
9. Kretschmer, K., Diehl, R., Hartmann, D.H., Line shape diagnostics of Galactic $^{26}$Al. *Astron. Astroph.*, **412**, L77-L81 (2003)
10. Knödlseder, J., Bennett, K., Bloemen H., *et al.*, A multiwavelength comparison of COMPTEL $^{26}$Al line data. *Astron. Astroph.*, **344**, 68-82 (1999)
11. Taylor, J.H., Cordes, J.M., Pulsar distances and the Galactic distribution of free electrons. *Astrophys. J.*, **411**, 674-684 (1993)
12. Cordes, J.M., Lazio, T.J.W., NE2001. I. A new model for the Galactic distribution of free electrons and its variability. *astro-ph*/**0207156** (2002)

13. Robin, A.C., Reylé, C., Derriere, R., Picaud, S., A synthetic view on structure and evolution of the Milky Way. *A&A*, **409**, 523-540 (2003)
14. Asplund, M, Grevesse, N. & Sauval, A. J. in: Cosmic Abundances as Records of Stellar Evolution and Nucleosynthesis, in honor of David L. Lambert, ASP Conference Series, **Vol. 336**, p.25 (2005)
15. Gilmore, G., The Star Formation History of the Milky Way. *ASP Conf. Proc.*, **230**, 3-12 (2001)
16. Boissier, S., & Prantzos, N., Chemo-spectrophotometric evolution of spiral galaxies – I. The model and the Milky Way. *Mon. Not. R. Astron. Soc.*, **307**, 857-876 (1999)
17. McKee, C.F., Williams, J.P., The luminosity function of OB associations in the Galaxy. *Astrophys. J.*, **476**, 144-165 (1997)
18. Reed, C.B., New estimates of the solar-neighborhood massive star birthrate and the Galactic supernova rate. *Astron.Journ.* **130**, 1652-1657 (2005)
19. Rauscher, T., Heger, A., Hoffman, R.D., Woosley S.E., Nucleosynthesis in massive stars with improved nuclear and stellar physics. *Astrophys. J.*, **576**, 323-348 (2002)
20. Limongi, M., & Chieffi, A., $^{26}$Al and $^{60}$Fe from massive stars. *Nucl.Phys.A,* **758**, 11c-14c (2005)
21. Palacios, A., Meynet, G., Vuissoz, C., *et al.*, New estimates of the contribution of Wolf Rayet stellar winds to the Galactic $^{26}$Al. *Astron. Astroph.*, **429**, 613-624 (2005)
22. Woosley, S. E., Heger, A., Hoffman, R. D., *in preparation for Astrophys. J.* (2005)
23. Kroupa, P., The initial mass function of stars. *Science*, **295**, 82-91 (2002)

# Radioactive $^{26}$Al and massive stars in the Galaxy

*Roland Diehl et al.*

## Online Supplementary Information Guide

1. Supplementary Methods: Observations and Data
   The INTEGRAL Observatory and its Spectrometer instrument have been launched in Oct 2002. Observations are composed of 7130 pointings along the plane of the Galaxy, and sum up to an exposure of 4 Ms at the Galactic Center, from the first two years of the mission.
   <Diehl_26Al-in-Galaxy_Nat_Suppl_1_Methods_ObsData.pdf><159kB>
2. Supplementary Methods: Data Analysis and Results
   Spectra are determined from independent model fits in 0.5 keV bins. Splitting the sky model into longitude segments allows for spatially-resolved spectroscopy, and obtains Doppler shifts as expected from Galactic rotation. The variability of resulting spectra with different models for the spatial distribution of $^{26}$Al emission is modest to small.
   <Diehl_26Al-in-Galaxy_Nat_Suppl_2_Methods_Gamma.pdf><212 kB>
3. Supplementary Discussion: Doppler Broadening
   The width of the observed gamma-ray line depends on the state of the ISM.
   <Diehl_26Al-in-Galaxy_Nat_Suppl_3_Discussion_Doppler.pdf><34 kB>
4. Supplementary Discussion: Galactic Rotation
   Different models for the spatial distribution of $^{26}$Al emission and rotation curves for the inner Galaxy lead to variations in expected line shifts.
   <Diehl_26Al-in-Galaxy_Nat_Suppl_4_Discussion_GalRot.pdf><66 kB>
5. Supplementary Discussion: Nucleosynthesis Yields
   Different models for stellar evolution and supernovae predict somewhat different yields of $^{26}$Al. From current models, an assessment is made over the full range of massive stars.
   <Diehl_26Al-in-Galaxy_Nat_Suppl_5_Discussion_Yields.pdf><82 kB>
6. Supplementary Methods: Deriving a Galactic Star Formation Rate from $^{26}$Al Gamma-rays
   The determination of the supernova rate follows from the nucleosynthesis yield and its integration over the mass distribution of stars. The conversion to a star formation rate is described.
   <Diehl_26Al-in-Galaxy_Nat_Suppl_6_Methods_SFR.pdf><92 kB>
7. Supplementary Discussion: Star Formation Rate (SFR) and Supernova Rate (SNR) Estimates for the Galaxy
   The different approaches determining supernova rates or star formation rates for the Galaxy are presented in a Table, with discussion of strengths and weaknesses. The $^{26}$Al-based approach is completely independent, and among the less-biased and more accurate methods.
   <Diehl_26Al-in-Galaxy_Nat_Suppl_7_Discussion_SFR.pdf><59 kB>

## Observations and Data

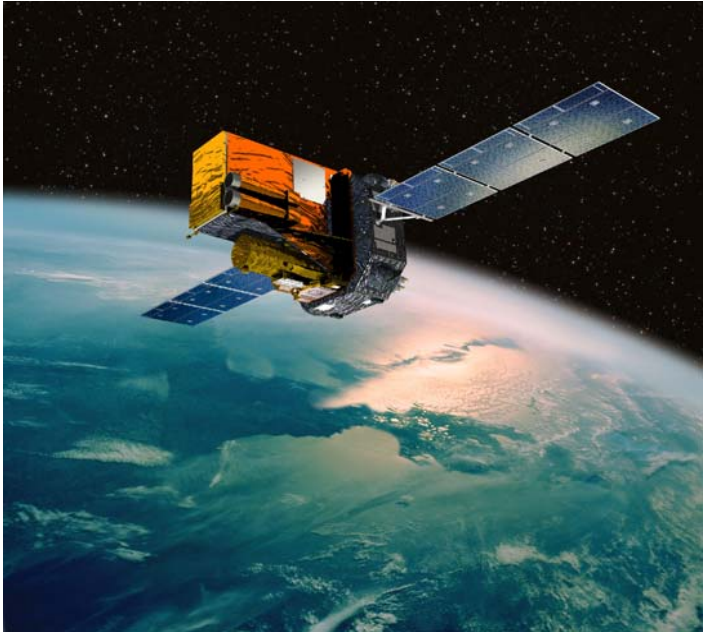

Fig. A1.1a: *ESA's INTEGRAL observatory*

The INTEGRAL[1] spacecraft of ESA, launched on Oct 17, 2002 from Baikonur, Kasachstan, carries two major instruments, both coded-mask telescopes for high-energy emission in the gamma-ray range (15 keV – 8 MeV): The "Imager" (IBIS)[2] features a finely-pixelised mask and a corresponding detector plane composed of 128x128 CdTe scintillation detectors, thus optimised for high imaging resolution up to 12 arcmin.

The "Spectrometer" (SPI)[3,4] is build around a 19-element camera of high-resolution gamma-ray detectors (detector size 72 x 55 x 55 mm), thus has coarser imaging capabilities with a resolution of 2.7°; its strength is an energy range covering gamma-ray lines from nuclear transitions, and a fine resolution of 3 keV at 1800 keV, which eases identification of the lines and opens the opportunity to even measure line shape details on individual lines, such as reported in this paper.

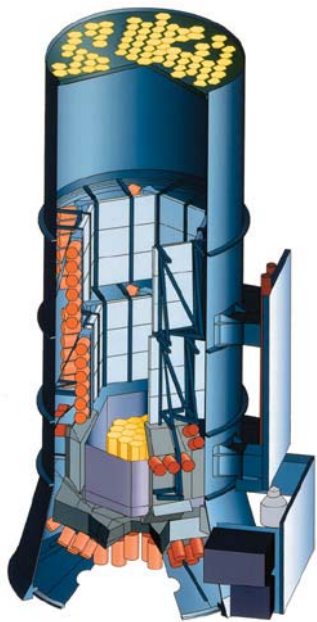

Fig. A1.1b: *The SPI Instrument*

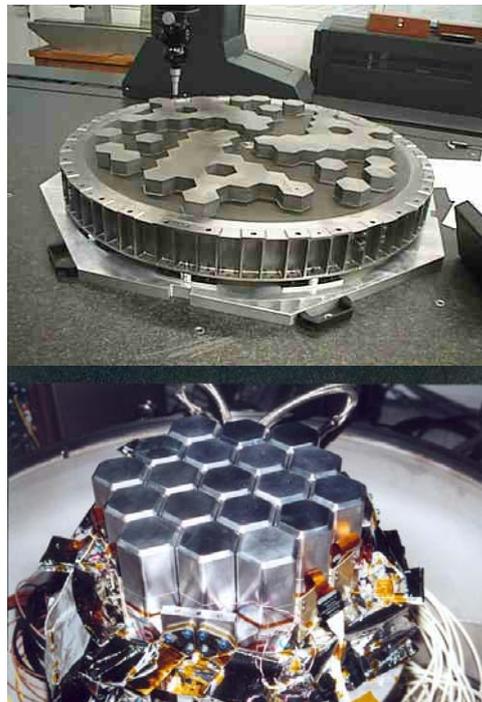

Fig. A1.1c: *SPI components: The Tungsten coded mask (left), and the 19-element Ge detector camera (right)*

Online Supporting Information: *Radioactive $^{26}$Al and massive stars in the Galaxy*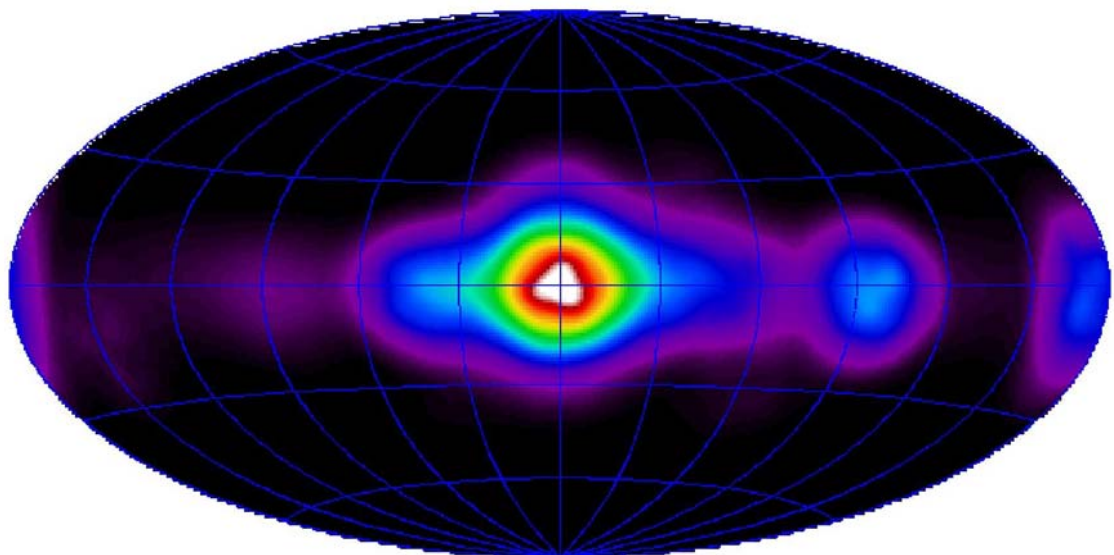

Fig. A1.2: *Effective exposure in Galactic coordinates (longitudes 180°…-180°, latitudes –90°…90°, 30° grid), with a linear colour scale. The exposure obtained at the Galactic Centre is 4 Ms*

The data analysed in this paper are assembled from observations during INTEGRAL orbits 15-259 (Nov 2002 – Nov 2004), using data from the Core Program and from observations of the Open Program which are either public or owned by members of our team. Altogether, this comprises an observation time of 16.5 Ms. We selected data to exclude solar flare periods and anomalies near orbital perigee from radiation belts. Our final dataset is composed of 7130 spacecraft pointings, each providing count spectra for each of the 19 detectors of the SPI instrument. Observations extend along the Galactic plane, yet concentrated towards the inner region of the Galaxy; secondary emphasis on the Vela, Crab, and Cygnus regions are evident.

**References**

1. Winkler, C., Courvoisier, T.J.-L., DiCocco, G., et al., The INTEGRAL mission. *A&A*, **411**, L1-L6 (2003)
2. Ubertini, P., Lebrun, F., DiCocco, G., et al., IBIS, the imager aboard INTEGRAL. *A&A*, **411**, L131-L140 (2003)
3. Vedrenne, G., Roques, J.-P., Schönfelder, V. et al., SPI, the spectrometer aboard INTEGRAL. *A&A*, **411**, L63-L70 (2003)
4. Roques, J.-P., Schanne, S., von Kienlin, A.. et al., SPI/INTEGRAL inflight performance. *A&A*, **411**, L91-L100 (2003)

## *Data Analysis and Results*
### Determination of $^{26}$Al Spectra.

SPI data are dominated by instrumental background, the celestial signal being on the order of 1-2%. We model background in our count spectra from independent tracers of cosmic-ray interactions with instrument and spacecraft. This background model is fitted to the measured count spectra, together with a spatial distribution model for celestial emission as convolved through the instrument response. We perform such a model fit in 0.5 keV wide energy bins across the region of the $^{26}$Al line, to obtain the intensity of the celestial emission model, together with several parameters for fine adjustment of the background model (Fig.1 of the main paper).

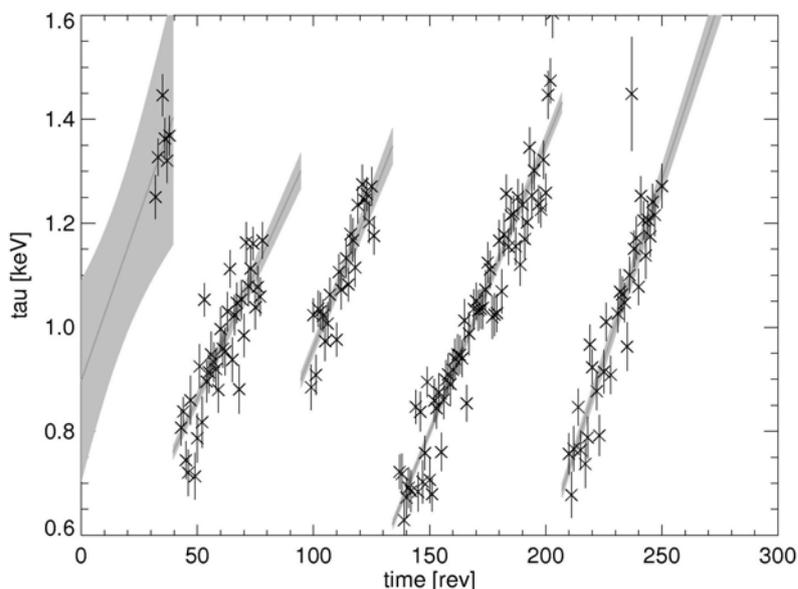

Fig. A2.1: *Time dependence of spectral resolution, as determined from fitting instrumental background lines. Shown is the degradation component τ of the spectral response; τ characterizes the width of a one-sided exponential to describe charge collection imperfections. Units of time are orbit revolutions ('rev', 3 days). Observations used in this paper cover orbits 15-259.*

The instrumental spectral response is affected by cosmic ray interactions in space, degrading the charge collection properties of the SPI Ge detectors (Fig.A.3). This degradation can be cured by heating of detectors to ~100°C for several days, an operation which has been performed successfully for SPI at intervals of typically 6 months. . We accumulate the effective instrumental line width from the model shown in Figure A2.1, and convolve a Gaussian line of parametrized width (for celestial $^{26}$Al) with this instrumental line shape to fit our spectrum (solid line in Figure 1 of main paper). This yields parameters for line intensity, centroid energy, and celestial line width; we include a linear function (2 parameters) to account for systematic or background effects which may produce an underlying bias across the $^{26}$Al line.

For determination of spectra for different sky regions (Fig. A2.2), we split the spatial distribution model into the sky regions of interest, and determine the sky brightness in each region and energy bin from the fit of the background and sky models simultaneously to our data. Here we adopt a symmetric source distribution in the inner Galaxy in the form of a double-exponential disk (scale radius 4 kpc, scale height 180 pc. (see Online Supplement 4 for estimates of systematics from different spatial distribution models). A symmetric model avoids bias from uncertainties of spatial distribution details, and is also plausible because spiral structure is insignificant in the inner Galaxy below ~35°). We split this model into three longitude intervals, to obtain spectra for these segments. For each of these spectra, we accumulate the effective instrumental line shape from the measurements that observed the respective region; convolving this shape with a Gaussian, we fit the line intensity, position, and width independently for each region. Statistics is reduced compared to Fig.1; therefore background in the spectra was assumed to be flat (as found in the total inner-Galaxy spectra, see Fig.1), to reduce the number of free parameters and avoid systematic distortions of line centroids. Using the central Galactic region (±10°) as a reference (fitted line energy 1808.70 keV), we find clear centroid energy shifts of -0.14 keV and +0.44 keV, respectively; expectations for this model are ±0.30 keV. The pattern of these shifts consistently appears for the plausible spatial models and variants of spectral fitting (see also Online Supplement 4).

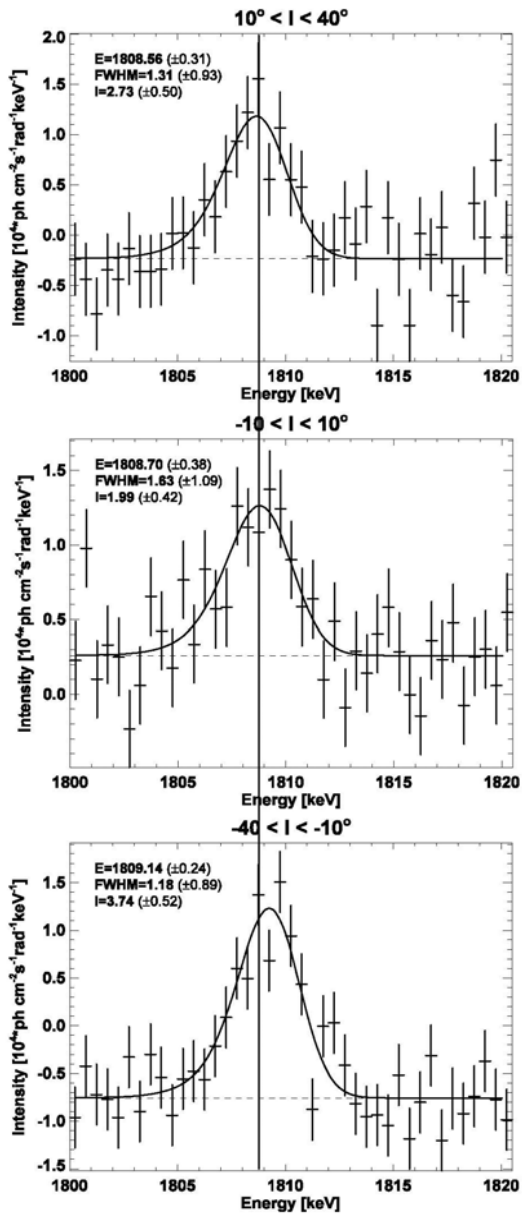

Fig. A2.2: *Measurements of the $^{26}$Al line from different longitude segments of the inner Galaxy region (s.d.error bars). Fits (solid line) implement the shape of the instrumental resolution as it results from cosmic-ray degradation and annealings during the time of measurements, convolved with a Gaussian for the intrinsic, astrophysical $^{26}$Al line width. Note that $^{26}$Al line emission is found to be brighter in the fourth quadrant, for this axisymmetric Galaxy model (see text), resulting in different error bars for line centroids in quadrants 1 and 4.*

**Determination of $^{26}$Al Flux.**
Due to the approach of model fitting per energy bin, resulting spectra may be expected to depend on the adopted model for spatial distribution of the $^{26}$Al emission. Models are based, as one family, on observations of $^{26}$Al directly with the COMPTEL instrument[1] and, alternatively, on plausible tracers of these sources. Candidate tracers[2] are warm dust emission seen in infrared light, or distributions of free electrons, because they reflect the energy inputs from massive stars, hence should correlate with the $^{26}$Al nucleosynthesis yields from these same massive stars. Another model family is derived through 3D geometrical distributions of sources in the Galaxy; these can be geometrical representations of the disk of the Galaxy[3], or deconvolved Galaxy components leading to dust emission[4], or decomposed models representing radio emission from electrons liberated by ionising radiation from massive stars[5,6]; here we only include components related to $^{26}$Al sources, and adjust the latitudinal scale height to what has been found for $^{26}$Al sources (~80-200 pc).

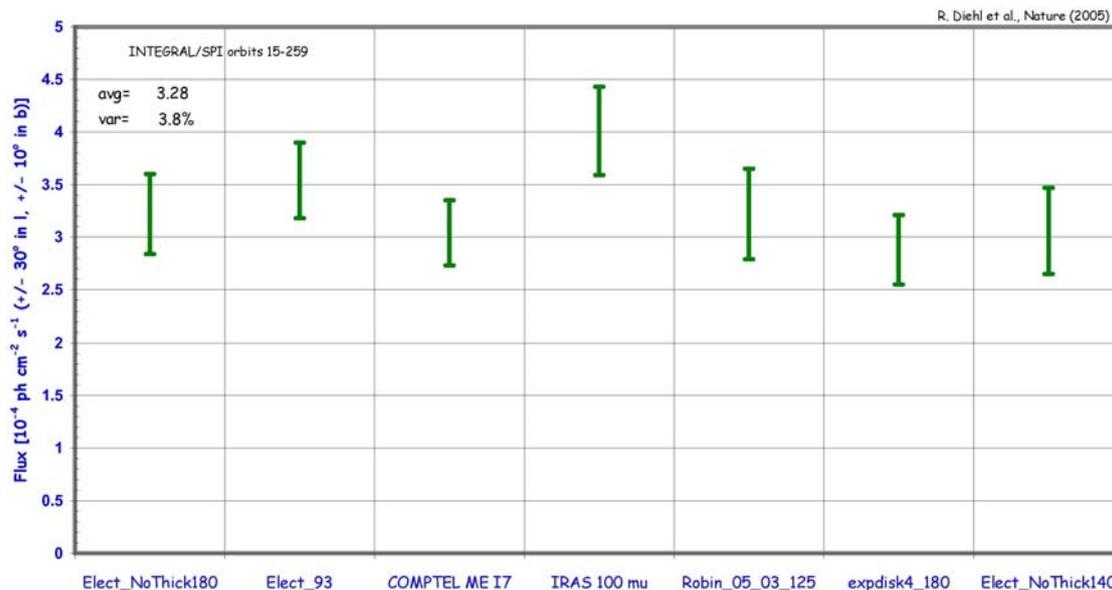

Fig. A2.3: *Derived $^{26}$Al intensity for different spatial-distribution models of $^{26}$Al emission. The COMPTEL Maximum-Entropy skymap[1] serves as a direct observational reference, while IR emission from warm dust had been found to be a rather good tracer of $^{26}$Al sources[2], and we use the IRAS map at 100 $\mu$m wavelength as a model. Emission from free electrons has been translated into different models[5,6] ('Elect…'); these encode the spiral structure of the Galaxy; different scale heights (140, 180 pc) were compared. Axisymmetric models may serve as reasonable representations of the $^{26}$Al source distribution, too; we use a simple double-exponential disk with Galactocentric scale radius 4 kpc and scale height 180 pc ('expdisk'), or, alternatively, a disk model which suppresses inner-Galaxy contributions in favor of a young stellar disk ('Robin…').*

Variations of the $^{26}$Al line parameters with spatial distribution models are rather small: The intrinsic $^{26}$Al line-width variance is 0.6%, hence our line width constraint is not affected. When we scan a family of plausible models, the inner-Galaxy $^{26}$Al flux variance is 4%, the total flux uncertainty is ~13% (with 12% from statistics, and ~4% estimated systematic uncertainty from the spatial distribution, added in quadrature) (see Figure A2.3).

For determination of line shapes for longitude segments, we reduce the free parameters of our spectral-shape fits by fixing offsets and slopes to zero, while using the instrumental line shapes as predicted for the exposure times of the respective longitude segments. This leaves as free parameters the line position, intensity, and intrinsic width, only.

**Determination of $^{26}$Al Mass.**

Flux values in the $^{26}$Al line as determined for geometrical source distribution models allow for a direct conversion of the measured gamma-ray flux into a total mass of radioactive $^{26}$Al contained in the Galaxy, because the 3D space distribution is known and can be integrated in terms of $^{26}$Al mass; normalization is obtained by standardizing the model projected sky brightness (we used the inner Galaxy region, -30°<l<30°; -10°<b<10°) and the brightness fit of the respective model to our data. From comparisons of a large family of models across all wavelength regions and plausible parameter ranges, respectively, models for the gas disk of the Galaxy, of warm dust emission based on the IRAS measurements, and free electrons as inferred from pulsar dispersion measurements, have been identified as most plausible[7-9,2].

For conversion to $^{26}$Al mass, we normalize the total integrated $^{26}$Al luminosity of our geometrical models to the line-of-sight integrated flux fitted to our data. Due to potentially very different distributions of $^{26}$Al in these models, the dispersion of derived $^{26}$Al amounts for the entire Galaxy is larger than the inner-Galaxy flux variations. Over the 4 models used (two exponential disk models, the IR dust geometrical representation, and the free electron based model), the average $^{26}$Al mass is 2.8 $M_\odot$ with a variance of 0.2 $M_\odot$. We conservatively estimate an uncertainty of ~30%, based on the $^{26}$Al flux and source distribution uncertainties in our analysis.

We use the mass of interstellar gas in the Galaxy[3] (4.95 $10^9$ $M_\odot$) and recently-updated standard abundances[10] (log $N_{Al}$=6.40 by number, normalized to log $N_H$=12) to determine the interstellar mass of Al, and thus obtain an isotopic ratio $^{26}$Al/$^{27}$Al of 8.4 $10^{-6}$

**References**

1. Plüschke, S., Diehl, R., Schönfelder, V., *et al.,* The COMPTEL 1.809 MeV Survey. *ESA-SP,* **459**, 55-58 (2001)
2. Knödlseder J., Bennett, K., Bloemen, H., et al., A multiwavelength comparison of COMPTEL 1.8 MeV line $^{26}$Al data. *A&A*, **344**, 68-82 (1999)
3. Robin, A.C., Reylé, C., Derriere, R., Picaud, S., A synthetic view on structure and evolution of the Milky Way. *A&A*, **409**, 523-540 (2003)
4. Drimmel, R., Galactic structure and radioactivity distributions. *New Ast. Rev.*, **46 (8-10)**, 585-588 (2002)
5. Taylor, J.H., Cordes, J.M., Pulsar distances and the Galactic distribution of free electrons. *ApJ*, **411**, 674-684 (1993)
6. Cordes, J.M., Lazio, T.J.W., NE2001. I. A new model for the Galactic distribution of free electrons and its variability. *astro-ph/***0207156** (2002)
7. Halloin et al., in preparation for *A&A* (2005)
8. Knödlseder J., Dixon, D., Bennett, K., et al., Image reconstruction of COMPTEL 1.8 MeV line $^{26}$Al data. *A&A*, **345**, 813-825 (1999)
9. Prantzos, N. & Diehl, R., Radioactive $^{26}$Al in the Galaxy: Observations versus theory. *Phys. Rep.,* **267, 1**,1 (1996)
10. Asplund, M, Grevesse, N., Sauval, A.J., The solar chemical composition. *ASP Conf. Series*, **CS336**, *in press*; (astro-ph/0410214) (2005)

## *Doppler Broadening and Interstellar Medium Dynamics*

If $^{26}$Al is mixed into hot interstellar medium, particle velocities due to thermal motion will cause a broadening of the emitted gamma-ray line. At high temperatures and low densities of the ISM, we may approximate this as an ideal gas, i.e. only the kinetic energy of the particles is important. Decaying $^{26}$Al nuclei are distributed in energy as $N(E) \propto \exp(-E/k_B T)$. Velocities are small compared to the speed of light, so only terms of first order in velocity *v* have to be considered, and it is sufficient to only consider the velocity component parallel to our line of sight. Therefore we observe a photon emitted with frequency $v$ by a particle moving with a velocity component $v_\parallel$ away from us at the Doppler shifted frequency $v' = v \cdot (1 - v_\parallel / c)$. As $E = 1/2 mv^2$, we obtain for the number of nuclei

$$N(v_\parallel) \propto \exp\left(-\frac{mv_\parallel^2}{2k_B T}\right)$$

which is a Gaussian with standard deviation $\sqrt{k_B T / m}$. Because of the linear relation between line-of-sight velocity and observed frequency, we see a line profile that is also Gaussian. Inserting the line energy *E* and atomic mass *m* of $^{26}$Al, we obtain the gas temperature corresponding to a measured line width:

$$T = \frac{(FWHM_E / E \cdot c)^2}{8 \ln 2 \, k_B} = 15.5 \cdot 10^6 \, \text{K} \left(\frac{FWHM_E}{1 \text{keV}}\right)^2 \left(\frac{m}{26 \text{u}}\right)$$

The root-mean-square average of the particle velocity in such a gas is

$$\sqrt{\overline{v^2}} = \sqrt{\frac{3 k_B T}{m}} = \frac{\sqrt{3} c \cdot FWHM_E}{2 \sqrt{2 \ln 2}} = 122 \, \text{km s}^{-1} \left(\frac{FWHM_E}{1 \text{keV}}\right)$$

i.e., typically a line width of 1 keV corresponds to velocities of 122 km s$^{-1}$.

For comparison with other physical origins of Doppler velocities such as interstellar turbulence or the expansion of a supernova remnant, the velocity distribution of nuclei must be treated accordingly. For example, in a homogeneously-expanding shell (e.g. a supernova explosion or a wind-blown cavity), the line shape can be derived by integration over the ring surface area $dA = 2\pi R \sin \alpha \, d\alpha$ on the sphere which the velocity vectors form in phase space ($\alpha$ being the angle between the line of sight and the particle velocity). The emission corresponding to d$\alpha$ spreads over

$$dE = E_0 \cdot \frac{dv_\parallel}{c} = \frac{E_0}{c} \cdot \frac{d}{d\alpha} |\mathbf{v}| \cos \alpha \, d\alpha \propto \sin \alpha \, d\alpha,$$

so becomes independent of $\alpha$ from dA, which results in a rectangular line profile. The velocity *v* as a function of the width $\Delta E$ is then

$$v = c \Delta E / 2 E_0 = 83 \, \text{km s}^{-1} (\Delta E / 1 \text{keV}).$$

An expansion velocity of 450 km s$^{-1}$ corresponds to a line width of 5.4 keV, which is the value quoted for the GRIS experiment[1]. This has been found difficult to understand in terms of our understanding of the ISM[2]. RHESSI and SPI measurements suggest a much lower line width around 1-2 keV[3-5].

[1] Naya, J.E., Barthelmy, S.D., Bartlett, L.M., *et al.*, Detection of high-velocity $^{26}$Al towards the Galactic centre. *Nature*, **384**, 44-46 (1996)

[2]. Chen, W., Diehl, R., Gehrels, N., *et al,.* Implications of the broad $^{26}$Al 1809 keV line observed by GRIS. *ESA-SP,* **382**, 105-108 (1997)

[3]. Smith, D., The RHESSI Observation of the 1809 keV line from Galactic $^{26}$Al. ApJ, **589**, L55 (2003)

[4]. Diehl, R., Kretschmer, K., Lichti, G.G., et al., SPI measurements of Galactic $^{26}$Al. A&A, **411**, L451 (2003)

[5]. Diehl, R., Halloin H., Kretschmer K., et al., $^{26}$Al in the inner Galaxy: Large-scale spectral characteristics derived with SPI/INTEGRAL. Submitted to A&A (2005)

## *Galactic Differential Rotation: Models and Observations of $^{26}$Al*

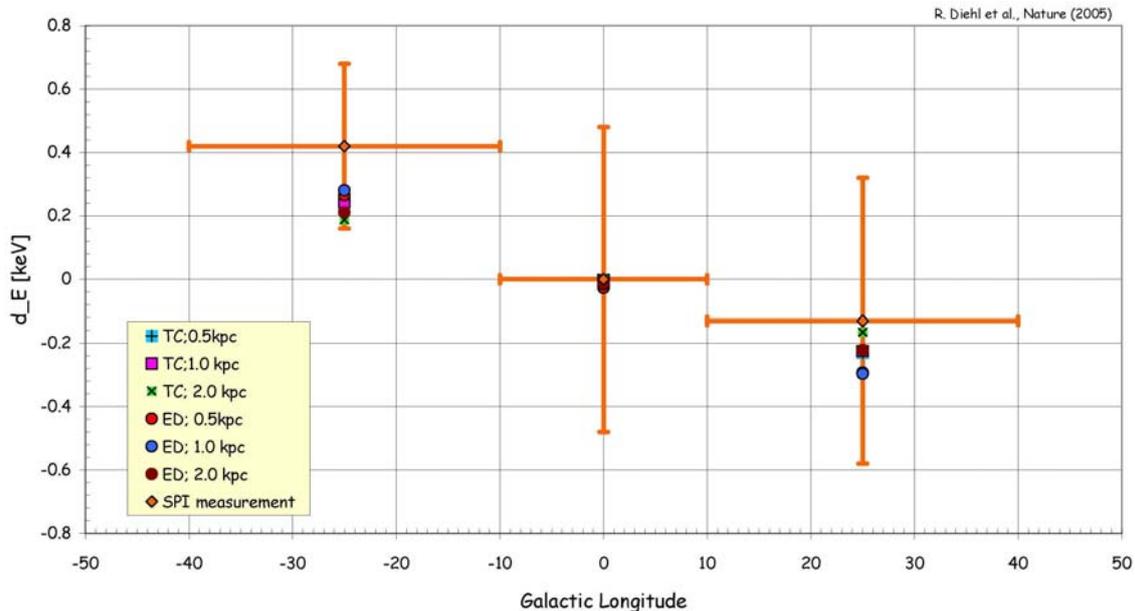

Fig. A4.1: Galactic rotation and $^{26}$Al Line Doppler Shifts: Here we compare, more quantitative and detailed, expectations from differential rotation of the inner Galaxy to our measurements. The rotation curve of the Galaxy is well-determined in the outer Galaxy regions only, extending inward to Galactocentric radii of ~2 kpc; nearer to the Galactic Centre, mass distribution models are used for extrapolation of the rotational behavior[1,2]. We conservatively estimate the impact of this uncertainty by variation of the inner-Galaxy slope of the rotation curves within a factor 2 of the suggested value (symbols). For a comparison to our measurements, which can only be obtained by summing the signal within a specified longitude segment, we also integrate a 3-dimensional $^{26}$Al source distribution model along the line-of-sight and within the same longitude segments. Our baseline $^{26}$Al source distribution model is derived from the free-electron determination by Taylor & Cordes (1993) based on pulsar dispersion measurements, which has been supplemented by adopting a latitudinal extent of the $^{26}$Al source distribution corresponding to a scale height above the disk of the Galaxy of 180 pc. Our measurements are fully consistent with these models ($\chi^2$ probability 0.94), suggesting that Galactic rotation is responsible for the observed small shifts of $^{26}$Al line energies with Galactic longitude.

*Annotations:*
- The numerical values and uncertainty estimates change slightly from the case of Fig 2 (in main paper), where a symmetric exponential disk with significantly larger contributions in the inner few 100 pc had been used.
- We again find the enhanced $^{26}$Al brightness in the fourth quadrant of the Galaxy with respect to the first quadrant (also seen in COMPTEL results), which is responsible for the larger error bar of the measurement here.

**References:**
1. Olling, R.P., Merrifield, M.R., *Luminous and dark matter in the Milky Way.* MNRAS, **326**, 164 (2001)
2. Olling, R.P., Merrifield, M.R., *Refining the Oort and Galactic constants.* MNRAS, **297**, 943 (1998)

## *Nucleosynthesis Yields*

Nucleosynthetic yields have been derived from models of nuclear burning in stars during their evolution and terminal core-collapse supernova. Stellar evolution includes core and shell hydrogen burning phases as main sources of $^{26}$Al, and in the supernova explosion, explosive nucleosynthesis occurs, mainly, in the O/Ne shell of the pre-supernova star. The complications of stellar structure and burning shells and their dependency on stellar masses leads to the variations of $^{26}$Al yields with mass, as they are reflected in current models (see solid red/green lines, the current best-evaluated models including both pre-supernova and explosive nucleosynthesis and mass loss). The comparison to models which address the stellar evolution of massive stars without the supernova ("Wolf-Rayet (WR) models") illustrates that explosive yields dominate up to at least 30 $M_\odot$; even at the highest masses (120 $M_\odot$) 50% of the ejected $^{26}$Al is produced explosively[2]. Furthermore, the different WR models[1,3,4] illustrate the impact of different treatments of mass loss and of stellar rotation in the models. Nuclear-reaction uncertainties are important, in particular for the neutron capture reactions destroying $^{26}$Al; these could change all these yields by factors of three. On the other hand, the treatment of stellar structure, evolutionary phases, and in particular convective mixing has been converging among different research groups, and yields agree within factors of two among different implementations of these aspects of nucleosynthesis. For reference, the often-used results by Woosley & Weaver (1995)[5] are shown as well, illustrating the effect of various advances over the last decade.

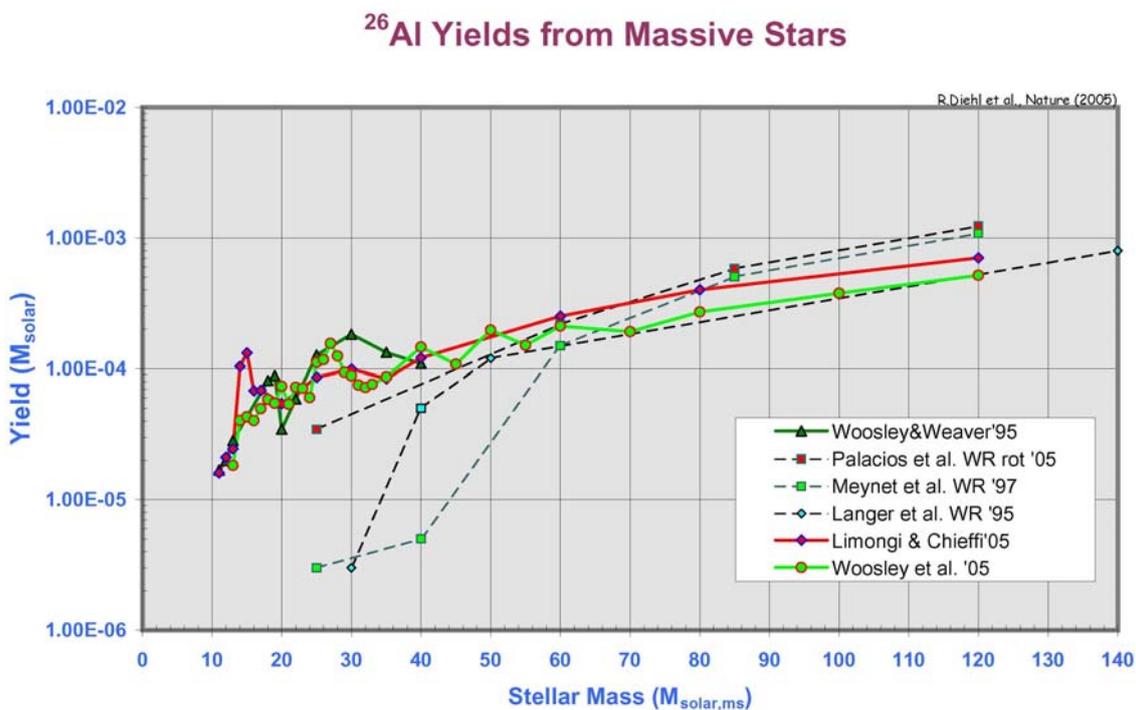

Fig. A5.1: *Nucleosynthesis yields of $^{26}$Al for massive stars.*

**References**
1. Langer N., Braun H., Fliegner J., The production of circumstellar $^{26}$Al by massive stars. Astr. & Sp.Sci., 224, 275-278 (1995)
2. Limongi M., Chieffi A., *in preparation for A&A.* (2005)
3. Meynet, G., Arnould, M., Prantzos, N., Paulus, G., Contribution of Wolf Rayet stars to the synthesis of $^{26}$Al. *A&A*, **320**, 460-468 (1997)
4. Palacios, A., Meynet, G., Vuissoz, C., *et al.*, New estimates of the contribution of Wolf Rayet stellar winds to the Galactic $^{26}$Al. *A&A*, **429**, 613-624 (2005)
5. Woosley, S.E., Weaver, T.A, The evolution and explosion of massive stars. *ApJS*, **101**, 181-235 (1995)
6. Woosley, S. E., Heger, A., Hoffman, R. D., *in preparation for ApJ.* (2005)

## *Deriving a Galactic Star Formation Rate from $^{26}$Al Gamma-rays*

Radioactive $^{26}$Al is produced and dispersed throughout the Galaxy at various sites of nucleosynthesis, and decays with a mean life of τ = 10$^6$ yrs. This process is dominated by ejected matter from core collapse supernovae and their preceding winds during the Wolf-Rayet stage. The relatively long mean life in comparison to the time between supernovae (~100 yrs) implies that a very large number of events contributes to a steady state abundance, which in turn results in a diffuse glow of the Galaxy in the 1.8 MeV gamma-ray line from the decay of $^{26}$Al. The basic idea of the "*$^{26}$Al method*" is the conversion of the observed total gamma-ray line flux for $^{26}$Al to the corresponding supernova rate (SNR), and consequently to the average star formation rate, SFR. The main advantage of this method is the lack of extinction corrections in the gamma-ray band, so that we can see the full disk of the Galaxy and are not limited to a small sampling volume around the Sun. It is also worthy to note that averaging over a few million years implies a large number of events, which provides a better statistical ensemble in comparison to other methods in which the tracers have a much shorter observational lifetime and often very significant selection effects and highly uncertain evolution corrections [pulsars, radio supernova remnants, historic SN record, HII regions].

The $^{26}$Al method relates the observed line flux to the steady state equilibrium mass of $^{26}$Al in the Galaxy via a spatial distribution model. Since the observed map at 1.8 MeV indicates that massive stars are the dominant contributors to this radioactive species in the ISM, we go beyond a simple axis-symmetric model, and include spiral structure of our Sbc-type Milky Way. The dynamic evolution of the injected $^{26}$Al should also be accounted for, as stellar winds and explosive outflows cause the density distribution to be more extended than what one would infer from the known spatial distribution of massive stars. The resulting density map is somewhat uncertain; but perhaps the largest source of potential error is the overall scale of the Galaxy: The official IAU distance of the Sun to the Galactic centre is 8.5 kpc, but a recent reviews of measurements[1,2] does indicate values as small as 7 kpc. Flux scales as the inverse square of distance, therefore such a global reduction in size of the Galaxy would decrease the inferred mass of $^{26}$Al by 34%.

The amount of $^{26}$Al is maintained in steady state by a core-collapse supernova rate (SNR) via $M_{eq} = SNR \cdot \tau \cdot Y$, where the rate is measured in events per year, τ is the mean life of $^{26}$Al, and *Y* is the IMF-averaged $^{26}$Al yield in units of M$_\odot$ per supernova. It must be emphasized that the yield in this context must include the explosive yields from the supernova model as well as any $^{26}$Al ejected in the Wolf-Rayet wind phase. Yields are moderated by the steep initial mass function (IMF), $\xi \sim m^{-\alpha}$, in our relevant mass range ~10-120 M$_\odot$. We use the Scalo IMF ($\xi \sim m^{-2.7}$) for this higher-mass range, supported by a wide range of astronomical constraints[6]. *Y* is obtained from the high-mass initial mass function (IMF) and the nucleosynthesis yields of models (see Online Supporting Information). The resulting $^{26}$Al yield per massive star is 1.4 10$^{-4}$ M$_\odot$, with an estimated uncertainty of 50% based on various published yields as a function of progenitor mass (see Online material on $^{26}$Al yields). The corresponding supernova rate is SNR = 1.92 events per century (this does not include supernovae of type Ia, which have been found to be negligible sources of $^{26}$Al), with an uncertainty of ~60% (from 30% due to Mass and 50% due to yield, added in quadrature). The resulting range of one to three core collapses per century coincides with the recent values obtained from on a survey of local O3-B2 dwarfs[3], extrapolated to the Galaxy as whole with spatial distribution models, and the study of the luminosity function of OB associations[9].

The conversion from core-collapse supernova rate, which assumes that all stars more massive than ten solar masses end their lives as supernovae[8], is given by $SFR = SNR \cdot <m> \cdot f_{SN}^{-1}$, where $<m>$ is the average stellar mass in a star formation event, and $f_{SN}$ is the fraction of all stars that become supernovae. Stars are predominantly formed in clusters with a quasi-universal IMF[6,7] (narrowly distributed about the canonical Salpeter power law), but that the integrated galaxial initial mass function (IGIMF) must be steeper than the canonical IMF. A steeper IGIMF implies a reduced supernova fraction in general, and implies a dependence on galaxy mass and the much less established cluster mass distribution function. It is thus possible that the true SFR could be significantly larger than the rate derived from the canonical IMF. We leave the application of such a correction to the reader, and report a SFR based on the conversion choice made by McKee and Williams[9]. With an average stellar mass of $<m>$ = 0.51 M$_\odot$ and $f_{SN}$ = 2.6 10$^{-3}$ we find SFR = (3.8 ± 2.2) M$_\odot$ yr$^{-1}$, which agrees with the value derived from the luminosity functions of OB associations[9], and implies a stellar production rate of $\dot{N}_* = SFR \cdot <m>^{-1}$ = 7.5 stars per year.

**References**

1. Reid, M. J. The Distance to the Center of the Galaxy, Ann. Rev. Astron. Astrophys. **31**, 345-372
2. Vallée, J.P., The spiral arms and interarm separation of the Milky Way: An updated statistical study. AJ, **130**, 569-575 (2005)
3. Reed, B. C. New Estimates of the Solar-Neighborhood Massive-Stars Birthrate and the Galactic Supernova Rate, submitted to AJ (see astro-ph/0506708) (2005)
4. Olling, R. P. & Merrifield, M. R. Two measures of the shape of the Milky Way's dark halo, MNRAS **311**, 361-369 (2001)
5. Timmes, F. X., Diehl, R., & Hartmann, D. H., Constraints from $^{26}$Al Measurements on the Galaxy's Recent Global Star Formation Rate and Core-Collapse Supernova Rate, ApJ, **479**, 760-763 (1997)
6. Kroupa, P. The Initial Mass Function of Stars: Evidence for Uniformity in Variable Systems, Science, **295**, 82-106 (2002)
7. Weidner, C. & Kroupa, P. The Variation of Integrated Star Initial Mass Functions among Galaxies, ApJ, **625**, 754-762 (2005)
8. Heger, A., et al. How Massive Single Stars End Their Lives, ApJ, **591**, 288-300 (2003)
9. McKee, C. F. & Williams, J. P. The Luminosity Function of OB Associations in the Galaxy, ApJ, **476**, 144-165 (1997)

# Star Formation Rate (SFR) and Supernova Rate (SNR) Estimates for the Galaxy

| Authors | SFR [$M_\odot y^{-1}$] | SNR [century$^{-1}$] | Comments |
|---|---|---|---|
| Smith et al. 1978 | **5.3** | 2.7 | |
| Talbot 1980 | **0.8** | 0.41 | |
| Guesten et al. 1982 | **13.0** | 6.6 | |
| Turner 1984 | **3.0** | 1.53 | |
| Mezger 1987 | **5.1** | 2.6 | |
| McKee 1989 | **3.6** (R) <br> **2.4** (IR) | 1.84 <br> 1.22 | |
| van den Bergh 1990 | 2.9 ± 1.5 | **1.5 ± 0.8** | „the best estimate" |
| van den Bergh & Tammann 1991 | 7.8 | **4** | extragalactic scaling |
| Radio Supernova Remnants | 6.5 ± 3.9 | **3.3 ± 2.0** | very unreliable |
| Historic Supernova Record | 11.4 ± 4.7 | **5.8 ± 2.4** | very unreliable |
| Cappellaro et al. 1993 | 2.7 ± 1.7 | **1.4 ± 0.9** | extragalactic scaling |
| van den Bergh & McClure 1994 | 4.9 ± 1.7 | **2.5 ± 0.9** | extragalactic scaling |
| Pagel 1994 | **6.0** | 3.1 | |
| McKee & Williams 1997 | **4.0** | 2.0 | used for calibration |
| Timmes, Diehl, Hartmann 1997 | 5.1 ± 4 | **2.6 ± 2.0** | based on $^{26}$Al method |
| Stahler & Palla 2004 | **4 ± 2** | 2 ± 1 | Textbook |
| Reed 2005 | 2-4 | **1-2** | |
| Diehl et al. 2005 | 3.8 ± 2.2 | **1.9 ± 1.1** | this work |

Table 1: *Star formation and core-collapse supernova rates from different methods.*

Generically, the SFR is obtained from a tracer that can be corrected for observational selection effects and is understood well enough so that possible evolutionary effects can be taken into account. One either deals with a class of residual objects, such as pulsars or supernova remnants, or with reprocessed light, such as free-free, H-alpha, or IR emission that follows from the ionization and heating of interstellar gas and its dust content in the vicinity of the hot and luminous stars. One must be careful to include time-dependent effects. The "after-glow" of an instantaneous starburst behaves differently than the steady-state output from a region with continuous star formation. We are concerned with an average star formation rate for the recent/current state of the Galaxy. We selected referenes for this table where either a supernova rate or a star formation rate is directly determined. The primary result in each paper is printed in **bold**.

Many papers discuss the star formation (rate) history (SFH) in relative terms, or the star formation rate surface density ($M_\odot y^{-1} kpc^{-2}$) in the solar neighborhood and its radial dependence. Such papers are not included in our table unless the global SFR is explicitly addressed or easily derivable.

We provide a *brief discussion of the various methods* used to estimate the Galaxy-wide star formation rate (in units of $M_\odot y^{-1}$) for the references quoted in our Table. Related quantities, such as production rate of stars (in units of stars $y^{-1}$) or the type-II (and Ib,c) supernova rate (in units of events century$^{-1}$) will also be quoted (if given). If only the SNR is given, we use the "calibration" from McKee & Williams (1997): SFR = 1.96 SNR, in the above units, resulting from <m> = 0.51 $M_\odot$ and $f_{SN}$ = 2.6 10$^{-3}$.

- Smith, L. F., Biermann, P., and Mezger, P. G. 1978 (A&A **66**, 65) find SFR = 5.3 $M_\odot$/yr.
- Talbot, R. J. 1980 (ApJ **235**, 821) finds SFR = 0.8 $M_\odot$/yr.
- Güsten, R. and Metzger, P. G. 1982 (Vistas in Astron. **26**, 3) find SFR = 13.0 $M_\odot$/yr. They attribute (5 ± 4) $M_\odot$/yr to spiral arm activity.
- Turner, B. E. 1984 (Vistas Astron. **27**, 303) finds SFR = 3.0 $M_\odot$/yr.
- Mezger, P. G. 1987 (in Starbursts and Galaxy Evolution, ed. T. X. Thuan, T. Montmerle & J. Tran Thanh Van, Paris, Editions Frontiers, p. 3) finds 5.1 $M_\odot$/yr.
- McKee, C. F. 1989 (ApJ **345**, 782-801) finds SFR = 3.6 $M_\odot$/yr from the analysis of thermal radio emission (free-free) from HII regions around massive stars. This emission is directly proportional to the production rate of ionizing photons, which in turn is directly proportional to the SFR. It is pointed out that this method is *very* sensitive to the slope of the high-mass IMF. It also must be noted that the method depends on stellar atmosphere models in conjunction of models for massive stars, which change with treatments of mass loss, rotation, and convection. This paper also briefly discusses the use of the far-IR luminosity, due to warm dust heated by the absorption of photons from massive stars. The author uses the measured IR luminosity of the Galaxy of 4.7 10$^9$ $L_\odot$ (from Mezger) to derive SFR = 2.4 $M_\odot y^{-1}$.
- Van den Bergh, S. 1990 (in "Supernovae", ed. S. E. Woosley, Springer Verlag, p. 711-719) finds (2.62 ± 0.8) $h_{100}^2$ century$^{-1}$. For $h_{100}$ = 0.75, the rate is 1.5 ± 0.8 century$^{-1}$. This rate is based on a combined study of galactic supernova remnants, historical SNe, and novae in M31 and M33. Cappellaro et al. refer to this rate as "the best estimate"
- Van den Bergh, S. and Tammann, G. 1991 (ARA&A **29**, 363-407) find SNR = 4.0 century$^{-1}$. The authors review supernova rates in external galaxies and derive a specific supernova frequency, in units of 1 SNu = one SN per century per 10$^{10}$ $L_\odot$(B), for various galaxy types. If one assumes that the Galaxy is intermediate between types Sab-Sb and types Sbc-Sd, the specific rate is ~3 $h_{100}^2$ SNu. For a Galactic blue-band luminosity of L(B) = 2.3 10$^{10}$ $L_\odot$(B) (their Table 11) and $h_{100}$ = 0.75 we infer SNR = 4.0 century$^{-1}$. This review paper also discusses estimates from Galaxy internal tracers: From radio supernova remnant (RSNR) statistics they find SNR = 3.3 ± 2.0 century$^{-1}$. From the historic record of nearby (< a few kpc) supernovae in the past millennium they (Tammann) find SNR = 5.8 ± 2.4 century$^{-1}$. The large extinction corrections in the galactic plane make this small sample highly incomplete, which results in very large uncertainties when one extrapolates to the full galactic disk. The authors also review efforts based on the pulsar birth rate, but their extensive observational selection effects in combination with strong (and poorly understood) evolution of luminosity and beaming geometry (see Lyne and Graham-Smith 1998, and Lorimer & Kramer 2005) renders this method impractical for estimating the galactic SNR.

- Cappellaro, E., et al. 1993 (A&A **273**, 383) find SNR = 1.4 ± 0.9 century$^{-1}$, based on scaling the rate in the Galaxy to that in external galaxies of similar type. The sample is obtained from surveys carried out at the Asiago and Sternberg Observatories. The authors provide an extensive discussion of the uncertainties of this method, which can exceed 200% for some late type galaxies.
- van den Bergh, S. & McClure, R. D. 1994 (ApJ **425**, 205-209) finds SNR = 2.4-2.7 century$^{-1}$, modulo $h_{75}^2$. This estimate is based on a reevaluation of the extra-galactic SN rates obtained from Evans's 1980-1988 observations. This method, as discussed above (van den Bergh and Tammann 1991) relies on an extrapolation from other galaxies, and thus a proper evaluation of the type of the Galaxy and its blue-band luminosity. The uncertainty due to the Hubble constant is now very small. Given the error analysis in the paper, the rate is uncertain by at least 34%. Thus we enter 2.5 ± 0.9 century$^{-1}$ into the table.
- Pagel, B. E. J. 1994 (in The Formation and Evolution of Galaxies, ed. C. Munoz-Tunon & F. Sanches, CUP, 110) finds SFR = 6.0 $M_\odot$/yr.
- McKee, C. F., Williams, J. P. 1997 (ApJ **476**, 144-165) find SFR = 4.0 $M_\odot$/yr, and based on the Scalo (1986) IMF convert this rate into a total number rate of 7.9 stars per year. They assume that all stars above 8 $M_\odot$ become supernovae, corresponding to a supernova fraction of $f_{SN}$ = 2.6 10$^{-3}$. The mean stellar mass is <m> = 0.51 $M_\odot$. No explicit uncertainty in the rate is provided. The corresponding cc-supernova rate is 2 century$^{-1}$.
- Timmes, F. X., Diehl, R., and Hartmann, D. H. 1997 (ApJ **479**, 760-763) find SFR = (5 ± 4) $M_\odot y^{-1}$. This method uses the same basic idea as presented in this work, but they utilize the $^{26}$Al line flux derived from the COMPTEL gamma-ray survey. The steady state equilibrium mass of $^{26}$Al obtained in their work was in the range 0.7 – 2.8 $M_\odot$, consistent, but lower than the value presented in this study. Based on the Salpeter IMF in the range 0.1 – 40 $M_\odot$ and the $^{26}$Al yields from Woosley & Weaver 1995 (ApJS **101**, 181) [which do not include contributions from the Wolf-Rayet wind phase] the authors derive the above quoted SFR and an associated cc-supernova rate of 3.4 ± 2.8 century$^{-1}$. The large uncertainty is mostly due to the equilibrium mass of $^{26}$Al inferred from the COMPTEL flux. INTEGRAL data presented here have significantly reduced the error in this key quantity.
- Stahler, S. W. & Palla, F. The Formation of Stars (Wiley-VCH) (2004) present (in their Chapter 19) the radial star formation rate density of the thin disk, based on HII regions, pulsars, and supernova remnants. With a solar neighbourhood calibration of 3 10$^{-9}$ M$_\odot$y$^{-1}$pc$^{-2}$ this integrates to about 4 M$_\odot$y$^{-1}$, and the authors suggest an error of at least 50% in this estimate.

- Reed, B. C. 2005 (AJ, in press – astro-ph/0506708) does not derive the SFR, but gives a statement on the SNR, "… the galactic supernova rate is estimated as probably not less than 1 nor more than 2 per century…". Using the conversion factors from McKee and Williams (1997), one could thus infer that the SFR is in the range 2-4 M$_\odot$y$^{-1}$. Reed uses a sample of a little over 400 O3-B2 dwarfs within 1.5 kpc of the Sun to determine the birthrate of stars more massive than 10 M$_\odot$. The Galaxy-wide rate is derived from this local measurement by extrapolation based on models for the spatial distribution of stars, a model for Galactic extinction (to accomplish corrections for stellar magnitudes), and a model of stellar life times. Reed emphasizes various sources of errors, such as lacking spectral classifications of some bright OB stars, the (unknown) inhomogeneous spatial structure of extinction as well as stellar density, and non-unique connection between mass and spectral type. Finally, Reed also draws attention to the fact that one would have to include B3 dwarfs as well, if the lower mass limit for supernovae is 8 M$_\odot$ and not 10 M$_\odot$ (as used in our study; see also Heger et al 2003). The OB-star catalog of the author was used to perform a modified V/V$_{max}$ test to obtain a present-day star count as a function of absolute V-band magnitude. From the stellar life times and the assumption of steady state the local birthrate follows. A double exponential model (in galactocentric radius and scale height above the plane) of the spatial distribution of these stars (which includes an inner hole of radius R = 4.25 kpc) ultimately leads to a total birthrate of 1.14 OB stars century$^{-1}$. Variations in the size of the hole change this number significantly, which leads the author to finally claim a rate of 1-2 supernovae century$^{-1}$.

*other references cited in the explanatory notes:*
- Heger, A., et. al. How Massive Single Stars End Their Lives. ApJ, **591**, 288-300 (2003)
- Lorimer, D. & Kramer, M. "Handbook of Pulsar Astronomy", CUP, (2005)
- Lyne, A. G., & Graham-Smith, F. "Pulsar Astronomy", 2$^{nd}$ ed., CUP (1998)